\documentclass[twocolumn]{aastex7}

\usepackage{xspace}
\usepackage[version=4]{mhchem}

\newcommand{\xsource}{4U 1820-30\xspace}

\begin{document}

\title{\textit A mysterious feature in the NICER spectrum of 4U 1820-30: A gravitationally redshifted absorption line?
}

\author[orcid=0000-0003-2882-0927, sname='Iaria']{R. Iaria}
\affiliation{Universit\`a  degli Studi di Palermo, Dipartimento di fisica e chimica - Emilio  Segr\`{e}, via Archirafi 36, Palermo, 90123,  Italia}
\email[show]{rosario.iaria@unipa.it}  

\author[orcid=0000-0002-3220-6375, sname='Di  Salvo']{T. Di Salvo} 
\affiliation{Universit\`a  degli Studi di Palermo, Dipartimento di fisica e chimica - Emilio  Segr\`{e}, via Archirafi 36, Palermo, 90123,  Italia}
\email{tiziana.disavo@unipa.itf}

\author[orcid=0000-0002-2701-2998, sname='Anitra']{A. Anitra} 
\affiliation{Universit\`a  degli Studi di Palermo, Dipartimento di fisica e chimica - Emilio  Segr\`{e}, via Archirafi 36, Palermo, 90123,  Italia}
\email{alessio.anitra@unipa.it}

\author[orcid=0000-0001-5852-6740, sname='Barra']{F. Barra} 
\affiliation{Universit\`a  degli Studi di Palermo, Dipartimento di fisica e chimica - Emilio  Segr\`{e}, via Archirafi 36, Palermo, 90123,  Italia}
\affiliation{Istituto di Astrofisica Spaziale e Fisica Cosmica, Istituto nazionale di astrofisica, via Ugo La Malfa 153, Palermo, I-90146, Italia}
\affiliation{Center for Astrophysics | Harvard \& Smithsonian, 60 Garden Street, Cambridge, MA 02138,USA}
\email{francesco.barra@unipa.it}

\author[orcid=0000-0002-0118-2649, sname='Sanna']{A. Sanna} 
\affiliation{Dipartimento di fisica, Universit\`a degli Studi di Cagliari, SP
Monserrato-Sestu, KM 0.7, Monserrato (CA), 09042, Italia}
\affiliation{Osservatorio Astronomico di Cagliari, Istituto nazionale di astrofisica, via della Scienza 5, Selargius (CA), I-09047, Italia}
\email{andrea.sanna@dsf.unica.it}

\author[orcid=0009-0007-8593-5006, sname='Maraventano']{C. Maraventano} 
\affiliation{Universit\`a  degli Studi di Palermo, Dipartimento di fisica e chimica - Emilio  Segr\`{e}, via Archirafi 36, Palermo, 90123,  Italia}
\affiliation{INAF – Osservatorio Astronomica di Brera, Via E. Bianchi 46, I-23807 Merate (LC), Italy}
\email{claudia.maraventano@unipa.it}

\author[orcid=0009-0005-1240-6985, sname='Miceli']{C. Miceli} 
\affiliation{Universit\`a  degli Studi di Palermo, Dipartimento di fisica e chimica - Emilio  Segr\`{e}, via Archirafi 36, Palermo, 90123,  Italia}
\affiliation{Istituto di Astrofisica Spaziale e Fisica Cosmica, Istituto nazionale di astrofisica, via Ugo La Malfa 153, Palermo, I-90146, Italia}
\affiliation{IRAP, Universitè de Toulouse, CNRS, UPS, CNES, 9, avenue du Colonel Roche, Toulouse, 44346 F-31028, France}
\email{carlotta.miceli@inaf.it}

\author[orcid=0000-0002-4773-3370, sname='Leone']{W. Leone} 
\affiliation{Universit\`a  degli Studi di Palermo, Dipartimento di fisica e chimica - Emilio  Segr\`{e}, via Archirafi 36, Palermo, 90123,  Italia}
\affiliation{Dipartimento di Fisica, University of Trento, Via Sommarive, 14, 38123 Povo (TN), Italy}
\affiliation{OATS, Via Giambattista Tiepolo, 11, 34131 Trieste (TS)}
\email{wladimiro.leone@unitn.it}

\author[orcid=0000-0001-5458-891X, sname='Burderi']{L. Burderi} 
\affiliation{Istituto di Astrofisica Spaziale e Fisica Cosmica, Istituto nazionale di astrofisica, via Ugo La Malfa 153, Palermo, I-90146, Italia}
\affiliation{Dipartimento di fisica, Universit\`a degli Studi di Cagliari, SP
Monserrato-Sestu, KM 0.7, Monserrato (CA), 09042, Italia}
\affiliation{Osservatorio Astronomico di Cagliari, Istituto nazionale di astrofisica, via della Scienza 5, Selargius (CA), I-09047, Italia}
\email{burderi@dsf.unica.it}

\begin{abstract}

A mysterious absorption feature at approximately 3.8 keV has been identified in the Neutron star Interior Composition Explorer (NICER) spectrum of the low-mass X-ray binary system 4U 1820-30.
We interpret this feature as a gravitationally redshifted iron absorption line. This interpretation is supported by the temporal proximity of the \textit{NICER} observation to the detection of a carbon superburst —a long and intense thermonuclear flash on the neutron star's surface— by the X-ray monitor MAXI, suggesting that the presence of the line is associated with this rare and extreme event. 
From the inferred redshift of the absorption line, the compactness of the neutron star can be derived. 
Fitting this feature with a photoionization absorption model, we measure a redshift of $1+z \simeq 1.72$, which implies a neutron star compactness of 
$R/M = 4.46\pm0.13$ km/M\(_{\odot}\) or $3.02\pm0.09$ in dimensionless units. 
This unique feature highlights the importance of further observations and detailed modelling, offering promising insights into the equation of state of matter under extreme density conditions.

\end{abstract}




\section{Introduction}

Neutron stars, remnants of core-collapse supernovae, serve as cosmic laboratories that provide a unique window into the physics of extremely dense matter. These stellar objects, with masses typically around 1.4 times that of the Sun but compressed into a sphere just about 10 kilometers in radius, exhibit conditions unattainable in terrestrial laboratories. The study of neutron stars enriches our understanding of stellar evolution and death and probes matter's fundamental properties under extreme pressures and densities.

\begin{figure*}[ht!]
\plotone{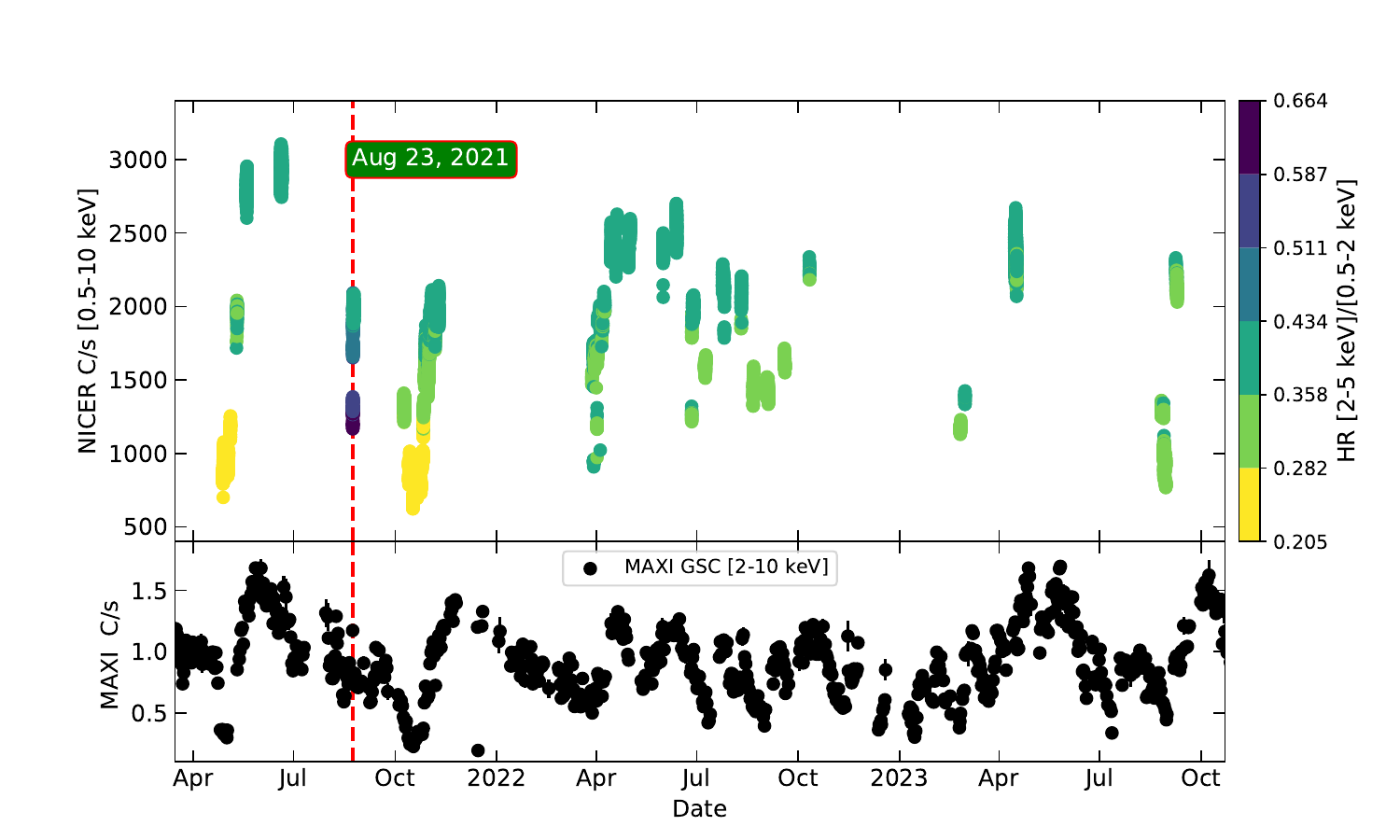}
\caption{
Comparative light curves of \xsource from \textit{NICER} (0.5–10 keV, top) and MAXI/GSC (2–10 keV, bottom). In the \textit{NICER} panel, color indicates hardness ratio (HR) between the [2–5] and [0.5–2] keV bands, ranging from 0.2 to 0.4. A red dashed line marks the observation with peak HR = 0.664. MAXI data show a brief increase in count rate coinciding with the \textit{NICER} observation window.
}
\label{fig:HR}
\end{figure*}

Central to the enigma of neutron stars is the equation of state (EoS) of ultra-dense matter, a pivotal yet elusive piece of the astrophysical puzzle. The EoS describes how matter behaves under the extreme conditions found within neutron stars, encompassing pressures and densities beyond nuclear saturation density. This quest to decipher the EoS challenges theoretical physicists and observational astrophysicists, requiring innovative theoretical models and cutting-edge observational techniques \citep{OzeL_2016ARA&A..54..401O}.

Recent years have witnessed significant advances in this domain, spurred by the advent of gravitational wave astronomy and high-precision X-ray observations. The detection of gravitational waves from neutron star mergers, such as GW170817, has provided new insights into the stiffness of the neutron star matter EoS and its behavior at supra-nuclear densities \citep{Abbott_2017PhysRevLett.119.161101}. 
Concurrently, observations from the Neutron star Interior Composition Explorer (\textit{NICER}) have offered new information on neutron star masses and radii, further constraining the possible EoS models \citep[see, e.g.,][]{Riley_2019ApJ...887L..21R}.

Among the most promising yet challenging methods to disclose information on the EoS of neutron stars is the detection of spectral lines from (close by) their surfaces. The gravitational redshift observed in these lines serves as a direct probe into the compactness of neutron stars.
The initial detection of gravitationally redshifted iron and oxygen absorption lines in the X-ray spectrum of the neutron star in EXO 0748-676 during type-I bursts marked a milestone in the study of ultra-dense matter 
\cite{Cottam_2002hrxs.confE..10C}. 
However, subsequent observations 
have yielded inconclusive results on the presence of the redshifted features initially reported \citep{Cottam_2008ApJ...672..504C}. Moreover, the observation of 
quasi-coherent oscillations during type-I bursts, indicative of the neutron star's spin at 552 Hz 
\citep{Galloway_2010ApJ...711L.148G}, 
cast doubts about the photospheric origin of the detected narrow absorption lines, 
that lack any Doppler broadening consistent with the fast spin.

During the 1980s, there were a few reports of X-ray absorption lines at 4.1 keV during type-I X-ray bursts observed with the Tenma and EXOSAT satellites \citep[see][]{Waki_1984PASJ...36..819W,Nakamura_1988PASJ...40..209N,Magnier_1989MNRAS.237..729M}.
These were interpreted as redshifted Ly$\alpha$ absorption from He-like iron at rest-frame energy of 6.7 keV. In all these cases, the observed absorption line energy of $\sim 4.1$ keV implies compactness of the neutron star of $R/M = 3.2$ (using $G=c=1$). Despite all these detections, the reality of these lines and their interpretation has been controversial, partly because subsequent observations have not confirmed these claims. 

Recent observational studies have revealed narrow emission and absorption features in Type-I bursts from 4U~1820$–$30, offering insight into the dynamics of photospheric radius expansion (PRE) events. \citet{2019ApJ...878L..27S} analyzed five PRE bursts observed with \textit{NICER} in 2017, identifying a discrete emission line near 1.0 keV and narrow absorption lines at $\sim$1.7 keV and $\sim$3.0 keV. These absorption features exhibited a systematic blueshift of $\sim$4.6\%, implying expansion velocities of order $v \approx 0.046c$ ($\sim$ 14,000 km s$^{-1}$), interpreted as due to burst-driven winds coupled with gravitational redshift effects \citep{2019ApJ...878L..27S}. Subsequently, \citet{Barra2025} extended the analysis to twelve PRE bursts, confirming up to four significant absorption lines per burst and consistently detecting an absorption feature at $\sim$2.97 keV. Together, these observations support a scenario in which heavy-element ashes are lifted into an expanding photospheric shell, producing discrete spectral lines whose blueshifts and strengths trace the dynamics and ionization structure of the burst-driven wind (see also \citealt{Zand_2010}, \citealt{2018ApJ...863...53Y}).

Recent analyses of thermonuclear X-ray bursts from the ultracompact binary 4U 1820–30 have revealed the presence of an absorption feature at approximately 3.8 keV. \cite{Peng2025} observed a redshifted absorption feature at $\sim 3.6 - 4$ keV in NICER data during the superburst recovery of 4U 1820–30. They attribute it to burning‑ash material in the accretion disk—likely Ar XVIII\footnote{It should be noted that the attribution by Peng et al.\ of the observed feature at 4.15~keV to Ar\,\textsc{XVIII} is inconsistent, as the rest-frame energy of the Ar\,\textsc{XVIII} Ly$\alpha$ transition is 3.32~keV. Instead, a transition at 4.1 keV would be consistent with the rest-frame energy of Car\,\textsc{XX} Ly$\alpha$.}—that becomes visible as the disk moves back inward.
A similar feature was independently reported by \cite{Jaisawal_2025}, who detected a 3.75 keV absorption line during the aftermath of a long burst event. In this case, the absorption line is attributed to absorption by heavy nuclei in the wind or accretion flow, enriched by nuclear burning products. The authors suggest that elements such as Si, Ar, Ca, or Ti may be responsible for this line, and that the absorption is shaped by both gravitational redshift and Doppler effects from the expanding photosphere.

However, the identification of an isolated absorption feature is intrinsically ambiguous when only a single line is detected. For this reason, here we fit the strong absorption line observed in the aftermath of the superburst from 4U 1820–30 with a self-consistent absorption model and find that it is well described by absorption from highly ionized, gravitationally redshifted iron, with a redshift factor of $1+z \approx 1.7$. The line is also significantly broadened, with $\Delta E_{\mathrm{FWHM}} / E \approx 0.15$), a width that is fully consistent with Doppler broadening due to the neutron star’s rotational velocity (assuming a spin frequency of 716 Hz) along the line of sight. We therefore interpret the line as originating from the atmosphere of the neutron star. The redshift of $1+z \simeq 1.72$ corresponds to a compactness of the neutron star of $\simeq 4.46$ km/M$_\odot$. If confirmed, this will offer invaluable insights into the EoS of matter at nuclear densities.

\section{Observations and analysis} \label{sec:data_analysis}

Following the detection of a bright X-ray burst from the position consistent with the low-mass X-ray binary 4U~1820$-$30 by the Monitor of All-Sky X-ray Image \citep[MAXI;][]{Serino_21} at 11:27~UT on 2021 August 23, \textit{NICER} initiated follow-up observations to monitor the post-burst evolution of the source. During the MAXI/GSC scan, the 4--10~keV flux, averaged under the assumption of constant emission, reached $1868^{+104}_{-94}$~mCrab (1$\sigma$ uncertainty). In the subsequent scan at 13:01~UT, the flux had decreased to $486 \pm 50$~mCrab, still above the persistent level of $\sim$200~mCrab. No significant excess was observed in the previous scan at 09:55~UT, suggesting that the burst onset occurred before the 11:27~UT transit.

\textit{NICER} \citep{Gendreau_16}  observed \xsource\ on 2021 August 23 at 14:35:00~UT (ObsID 4050300105, hereafter Obs.~105) and again on August 24 at 06:08:25~UT (ObsID 4050300106, hereafter Obs.~106), with exposures of 25~ks and 52~ks, respectively. The X-ray Timing Instrument (XTI) aboard \textit{NICER} operates in the 0.2--12~keV range and consists of 56 photon detectors, of which 52 Focal Plane Modules (FPMs) are currently active; four modules (12, 22, 29, and 60) were disabled post-launch due to damage.

Data reduction was carried out using the \texttt{nicerl2} pipeline within \texttt{NICERDAS} v13, part of the HEASOFT v6.34 package \citep{Blackburn_1995}, using standard filtering criteria and calibration database version \texttt{xti20240206}. All operational FPMs were active during both observations. Light curves in the 0.5--2~keV and 2--5~keV bands were extracted with \texttt{nicerl3-lc}, and corresponding hardness ratios (HRs) were computed.

Figure~\ref{fig:HR} displays the light curves and HRs for Obs.~105 and Obs.~106. During Obs.~106, the count rate in the 0.5--2~keV band increased from 700~cts/s to 1400~cts/s, while the 2--5~keV band rose from 450~cts/s to 600~cts/s, before decreasing to $\sim$525~cts/s at 110~ks from the start. The HR initially dropped steeply from 0.66 to 0.55 in the first 3~ks, followed by a more gradual decline to 0.4.

For Obs. 105 and  Obs. 106, light curves from two distinct energy bands and their corresponding HRs are presented in Figure \ref{fig:HR}. The analysis revealed a count rate increase in the 0.5-2 keV band from 700 c/s to 1.400 c/s and in the 2-5 keV band from 450 c/s to 600 c/s, before a decrease to 525 c/s at 110~ks from the start. The HR showed a steep decline from 0.66 to 0.55 in the first 3~ks, followed by a more gradual decrease to 0.4.

\begin{figure}[ht!]
\centering
\includegraphics[width=.5\textwidth]{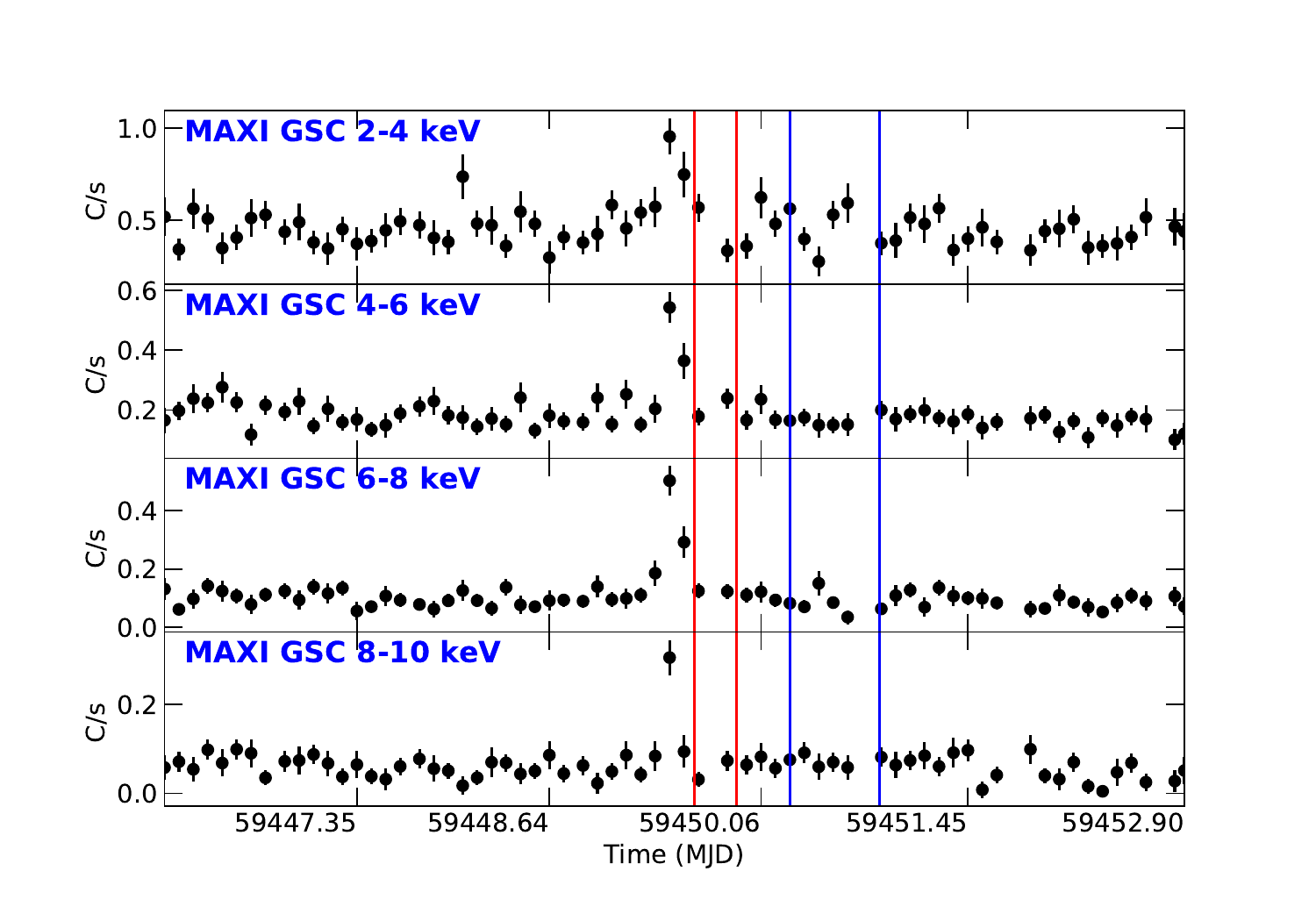}
\caption{Light curves of the source \xsource observed by MAXI in four different energy bands from MJD 59446 to MJD 59453. The vertical lines in red and blue mark the commencement and conclusion of \textit{NICER} Obs. 105 and 106, respectively. The noticeable surge in the count rate across all selected energy bands correlates with a superburst event, which occurred approximately six hours prior to the initiation of \textit{NICER} Obs. 105. }
\label{fig:maxi}
\end{figure}

The MAXI light curves of \xsource, obtained with the Gas Slit Camera (GSC) in four energy bands from MJD 59446 to MJD 59453 and extracted adopting the method described in \cite{Nakahira_12}, are showcased in \autoref{fig:maxi}. The red and blue vertical lines demarcate the start and end of \textit{NICER}'s Obs. 105, highlighting the superburst's onset approximately six hours before Obs. 105 began. This event's impact is notably pronounced in the 4-6 keV, 6-8 keV, and 8-10 keV bands, where the source's intrinsic variability is less dominant.
The 2-20 keV MAXI count rate during the NICER observations is $0.92\pm0.03$ c/s.  
Looking at the Hardness-Intensity diagram (HID) of \xsource presented by \cite{Marino23}, a count rate of $0.92\pm0.03$ c/s suggests that the source is in the Banana state \cite[see e.g.,][]{2023hxga.book..147D}.

Initially, we investigated the averaged spectrum obtained from Obs. 105 employing the \texttt{nicerl3-spect} tool and excluding Focal Plane Modules (FPMs) 14 and 34 due to their high detector noise levels. This selection criterion prevented possible distortions of the spectral data. The extraction of source spectra and corresponding {\tt scorpeon} background files\footnote{\texttt{scorpeon} background model in XSPEC: \url{https://heasarc.gsfc.nasa.gov/docs/nicer/analysis_threads/scorpeon-xspec/}} was performed with \texttt{nicerl3-spect}, utilizing the \texttt{bkgformat=file} option. In alignment with the \textit{NICER} calibration team's recommendations for systematic error incorporation\footnote{\textit{NICER} calibration guidelines: \url{https://heasarc.gsfc.nasa.gov/docs/nicer/analysis_threads/cal-recommend/}}, we 
added a systematic error to all the extracted spectra.
In particular,
by setting \texttt{syserrfile=CALDB} in \texttt{nicerl3-spect}, we introduced a systematic error of 1.5\% across the 0.4-10 keV energy band.
 The data were then optimally binned using the {\tt ftgrouppha} tool, applying the \texttt{grouptype=optmin} strategy to ensure a minimum of 25 counts per energy bin, as advised by \cite{Kaastra_16} for statistical robustness in spectral fitting. The exposure time of the spectrum is 8 ks. 

\subsection{The Averaged Spectrum from Obs. 105}
\begin{deluxetable*}{l@{\hskip 2pt}l@{\hskip 2pt}c@{\hskip 3pt}c@{\hskip 3pt}c@{\hskip 3pt}c}
\tablecaption{Best-fit values of the averaged spectrum obtained from Obs. 105. \label{tab:parameters_averaged_105}}
\tablewidth{0pt}
\tablehead{
\colhead{Component} & \colhead{Parameter} & \colhead{Model 1} & \colhead{Model 2} & \colhead{Model 3} & \colhead{Model 4}
}
\startdata
\texttt{TBfeo} & $N_{\rm H}$ & $0.172\pm0.003$ & $0.177\pm0.005$ & $0.192\pm0.009$ & $0.190^{+0.008}_{-0.006}$ \\
 & $A_{\rm O}/A_{\rm O}^{\odot}$ & $1.16\pm0.07$ & $1.19\pm0.07$ & $1.24\pm0.07$ & $1.27\pm0.08$ \\
\texttt{diskbb} & $kT_{\rm disk}$ (keV) & $0.70\pm0.05$ & $0.58\pm0.04$ & $0.48\pm0.05$ & $0.53\pm0.05$ \\
 & $R_{\rm disk}\sqrt{\cos{i}}$ (km) & $16\pm2$ & $22\pm3$ & $31\pm6$ & $27\pm4$ \\
\texttt{swind1} & $N_{{\rm H,wind}}$ & -- & -- & -- & $4^{+3}_{-1}$ \\
 & $\log(\xi)$ & -- & -- & -- & $3.24^{+0.33}_{-0.13}$ \\
 & $\sigma$ ($v/c$) & -- & -- & -- & $0.06\pm0.02$ \\
 & $z$ & -- & -- & -- & $0.70^{+0.03}_{-0.02}$ \\
\texttt{gaussian} & $E$ (keV) & -- & $3.80\pm0.04$ & $3.80\pm0.05$ & -- \\
 & $\sigma$ (keV) & -- & $0.24\pm0.04$ & $0.25\pm0.05$ & -- \\
 & $I$ ($\times 10^{-2}$) & -- & $-0.75\pm0.15$ & $-0.7\pm0.2$ & -- \\
\texttt{thcomp} & $\Gamma$ & $3.3^{+0.4}_{-0.2}$ & $2.82^{+0.15}_{-0.11}$ & $2.73^{+0.13}_{-0.10}$ & $3.02\pm0.13$ \\
 & $kT_e$ (keV) & (3.1) & (3.1) & (3.1) & (3.1) \\
\texttt{bbodyrad} & $kT_{\rm bb}$ (keV) & $1.23\pm0.07$ & $1.08\pm0.05$ & $1.04\pm0.04$ & $1.12\pm0.04$ \\
 & $R_{\rm bb}$ (km) & $11.7\pm1.2$ & $15.2\pm1.3$ & $16.5\pm1.5$ & $14.3^{+1.2}_{-0.8}$ \\
\texttt{gaussian} & $I_{\text{\ion{Mg}{12}}}$   ($\times 10^{-2}$) & -- & -- & $1.0\pm0.6$ & $1.6^{+0.4}_{-0.6}$ \\
\texttt{gaussian} & $I_{\text{\ion{Si}{14}}}$  ($\times 10^{-2}$) & -- & -- & $1.3\pm0.5$ & $1.8\pm0.4$ \\
\texttt{gaussian} & $I_{\text{\ion{S}{15}}}$  ($\times 10^{-2}$) & -- & -- & $0.4\pm0.3$ & $0.9\pm0.3$ \\
\texttt{gaussian} & $I_{\text{\ion{Ar}{17}}}$  ($\times 10^{-2}$) & -- & -- & $0.3\pm0.2$ & $0.6\pm0.3$ \\
$\chi^2$ (dof) & & $233.1\,(163)$ & $109.9\,(160)$ & $81.3\,(156)$ & $97.6\,(155)$ \\
\enddata
\tablecomments{The equivalent hydrogen column densities are in units of $10^{22}$ cm$^{-2}$. Errors are reported at a 90\% confidence level. Values in round parentheses were kept fixed during the fit. The energies and widths of the emission lines were fixed to the values reported in the text.
}
\label{tab:averagespectrum}
\end{deluxetable*}
\begin{figure*}[ht!]
\centering
\includegraphics[width=.4\textwidth]{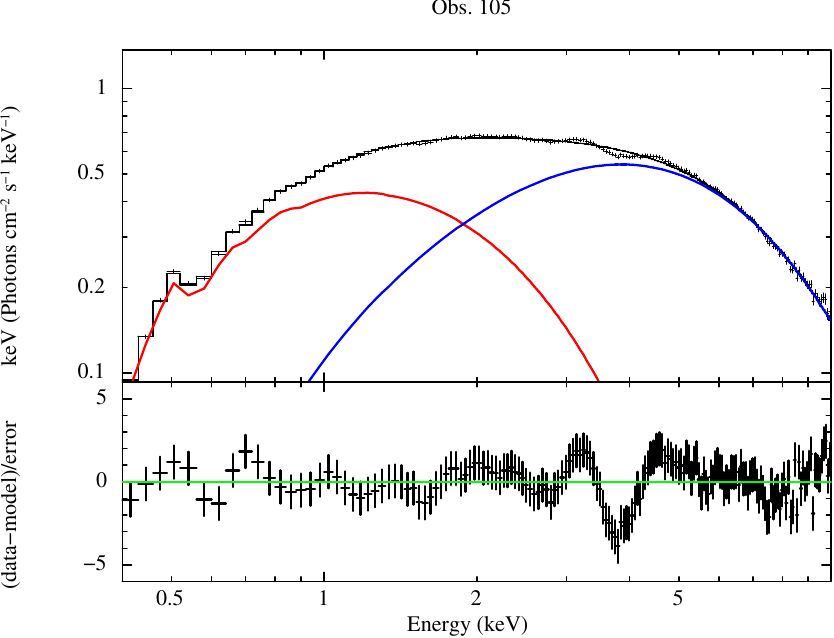} 
\includegraphics[width=.44 \textwidth]{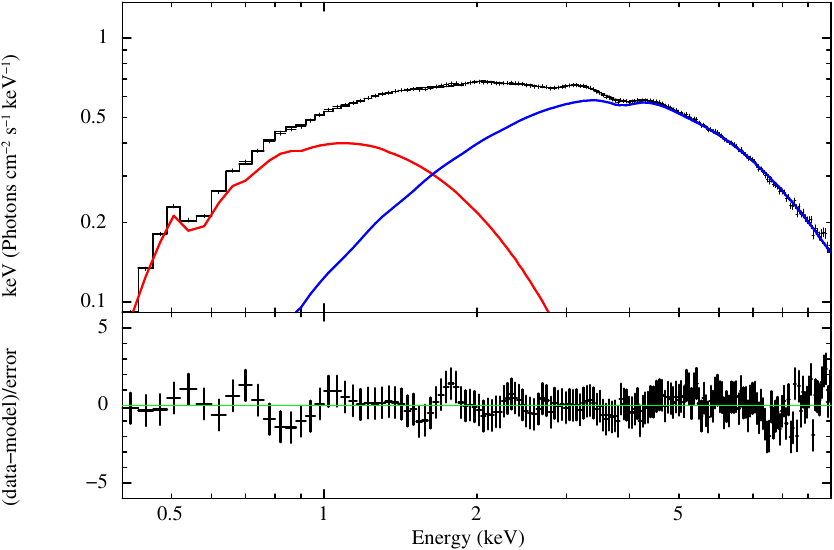}
\caption{Unfolded spectrum and residuals in unit of $\sigma$ for the averaged spectrum obtained from the   \textit{NICER} Obs. 105, adopting  \texttt{Model 1} (left panel) and \texttt{Model 4} (right panel). 
    The thermal emission from the accretion disk and the Comptonized component are in red and blue, respectively. }
\label{fig:pl_euf_4parti}
\end{figure*}
 \begin{figure}[ht!]
\centering
\includegraphics[width=.5\textwidth]{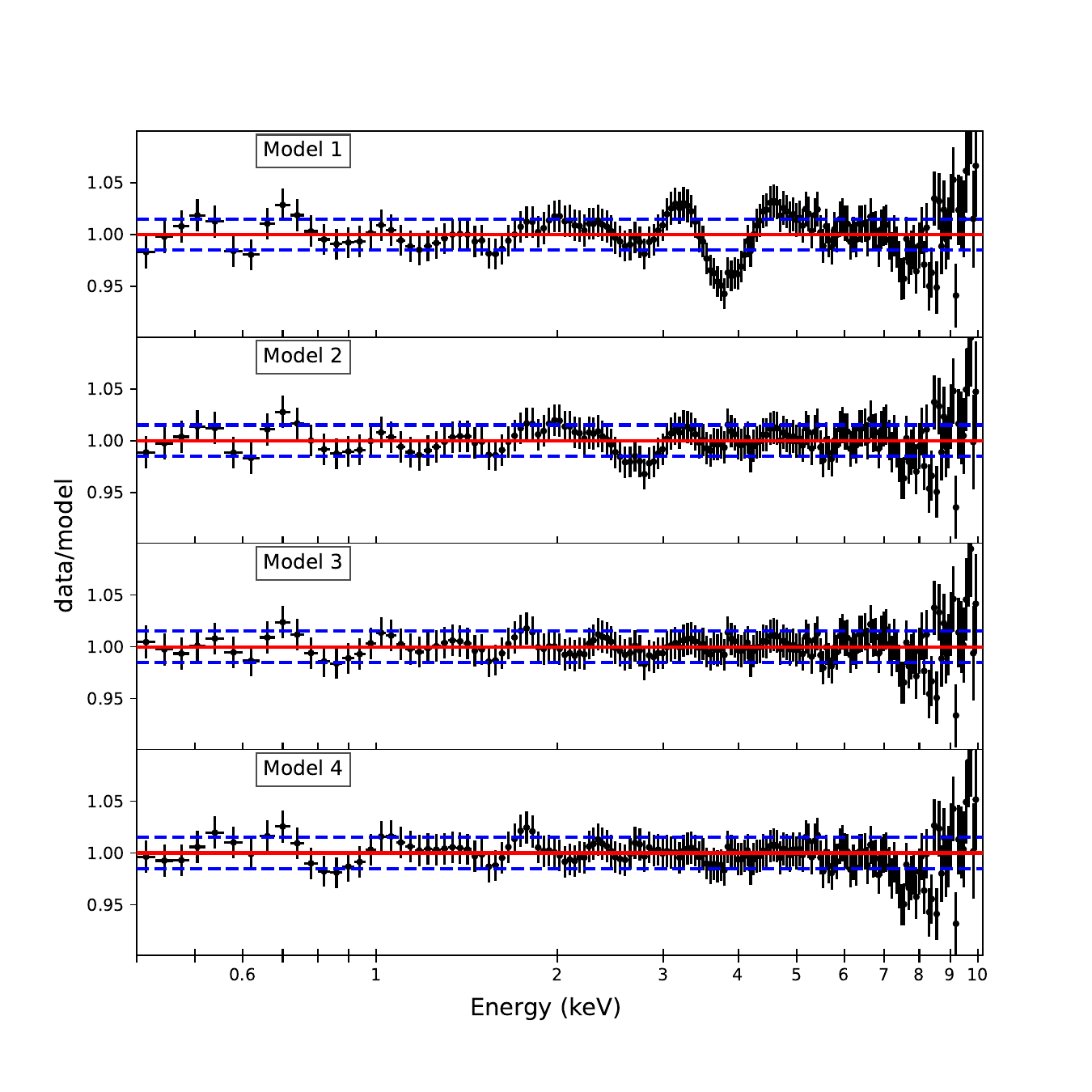}
\caption{Data-to-model ratios of the averaged spectrum extracted from Obs. 105, adopting \texttt{Model 1}, \texttt{Model 2}, \texttt{Model 3}, and \texttt{Model 4} (from top to bottom, respectively). The dashed blue lines indicate a ratio of 1.5\%, corresponding to the systematic error associated with the spectrum. }
\label{fig:pl_ratio_tot_rev}
\end{figure}
We perform spectral fitting across the 0.4--10~keV energy range of the averaged spectrum obtained from the \textit{NICER} Obs. 105, using XSPEC v12.14.1 and adopting the interstellar abundances of \cite{Wilms00} and the photoelectric cross sections of \cite{Verner_96}.
   The baseline model (Model~1) was defined as \texttt{TBfeo*(diskbb+thcomp*bbodyrad)}. The \texttt{TBfeo} component models interstellar absorption, allowing oxygen and iron abundances to vary. An excess at 0.54--0.56~keV was modeled as an overabundance of neutral oxygen, consistent with previous studies of 4U~1820--30 \citep{Costantini_12}.

The \texttt{diskbb} component describes multitemperature disk blackbody emission, while \texttt{thcomp*bbodyrad} represents Comptonized blackbody emission \citep[see][for a detailed description of the convolutive component \texttt{thcomp}]{2020MNRAS.492.5234Z}, assuming full coverage (cov = 1) and a fixed electron temperature of 3.1~keV, based on prior broadband studies \citep{Marino23}. The blackbody seed temperature, spectral index $\Gamma$, and normalization were left free to vary.

This model yielded a reduced chi-square of 1.43 for 163 degrees of freedom. Examination of the residuals revealed an absorption feature near 3.8~keV. We first modeled this using a negative Gaussian component (Model~2, \texttt{TBfeo*(diskbb+Abs$_{\text{line}}$+thcomp*bbodyrad)}), which significantly improved the fit. Further refinement included four Gaussian emission lines at fixed energies (identified with transitions of \ion{Mg}{12}, \ion{Si}{14}, \ion{S}{15}, and \ion{Ar}{17}, respectively) to account for low-energy residuals (Model~3, \texttt{TBfeo*(diskbb+Abs$_{\text{line}}$+4Em$_{\text{line}}$+thcomp*bbodyrad)}). 
The energies of four emission lines were fixed at their respective rest-frame values: 1.472~keV, 2.006~keV, 2.46~keV, and 3.13~keV. The line widths were fixed at 0.2~keV. The origin of these broad emission lines is consistent with reflection from the inner regions of the accretion disk, where relativistic effects and ionized disk material can produce broadened features at these energies \citep[see][]{Anitra2025}.

To evaluate the significance of the broad absorption line, we determined the number of standard deviations by which its normalization deviates from zero. This analysis yielded a significance of $6.1\sigma$. The energy of the absorption line is $3.80 \pm 0.05$ keV, the width is $0.25 \pm 0.05$ keV and the normalization is $(-0.7\pm 0.2) \times 10^{-2}$ photons cm$^{-2}$ s$^{-1}$. 
The equivalent width of the line was measured to be $44 \pm 7 \, \mathrm{eV}$, with its uncertainty reported at a 68\% confidence level.

In order to identify the 3.8-keV absorption line, we substituted the Gaussian line with a physically motivated model incorporating the \texttt{swind1} component \citep{2006MNRAS.371L..16G} to represent partially ionized absorption with velocity shear (Model~4, \texttt{TBfeo*(diskbb+4Em$_{\text{line}}$+swind1*thcomp*bbodyrad)}).
 {\tt  swind1} describes a partially ionized absorbing material with large velocity shear, approximated using {\tt XSTAR} kn5 photoionization absorption model grids (calculated assuming a microturbulent velocity of 100 km/s) and then convoluted with Gaussian smearing. {\tt  swind1} is defined by four parameters: the ionization state of the absorber ($\xi = L / nR^2$, where $L$, $n$, and $R$ are the X-ray luminosity, particle density, and distance of the cloud from the ionizing source, respectively), which can change in response to the variability of the illuminating continuum; the hydrogen column density ($N_{H_{\text{wind}}}$), which can change as clouds move in and out of the line of sight; the velocity shear, parameterized by the width of the smearing Gaussian ($\sigma$ in units of v/c), which can also change; and finally, the redshift ($z$).

This model yielded a best-fit $\chi^2$ of 97.6 for 155 degrees of freedom. The inclusion of the {\tt  swind1} component successfully accounted for the broad absorption feature without the need for an \textit{ad hoc} Gaussian and without leaving any residuals that might indicate model-predicted features not present in the observed spectrum. Furthermore, the multiplicative {\tt  swind1} component was applied to the Comptonized emission in order to mimic the presence of an ionized absorber located near the neutron star surface.

The observed disk and blackbody temperatures, \(kT_{\mathrm{disk}}\) and \(kT_{\mathrm{bb}}\), are \(0.53 \pm 0.05\) keV and \(1.12\pm0.04\) keV, respectively. Using the obtained blackbody normalization values and assuming a distance of \(8.0 \pm 0.3\) kpc to \xsource's host, the globular cluster NGC 6624 \citep{Baumgardt_21}, we calculated the inner radius of the accretion disk, uncorrected for the inclination angle \(i\), as \(R_{\mathrm{disk}}\sqrt{\cos{i}} = 27\pm4\) km. Similarly, the blackbody radius \(R_{\mathrm{bb}}\) was determined to be \(14.3^{+1.2}_{-0.8}\) km.

The large blackbody radius suggests that the seed photons for the Comptonized cloud originate from the neutron star surface and the boundary layer.
The power-law photon index of the Comptonization is \(\Gamma = 3.02\pm0.13\). Having fixed the electron temperature of the Comptonizing cloud at 3.1 keV, this corresponds to an optical depth \(\tau = 6.0\pm0.4\).

Finally, the equivalent hydrogen column density of the ionized absorbing matter responsible for the broad absorption line observed in the spectrum is \(N_{H_{\mathrm{wind}}} = (4^{+3}_{-1}) \times 10^{22} \, \mathrm{cm}^{-2}\). The ionization parameter is \(\log(\xi) = 3.24^{+0.33}_{-0.13}\), and the Gaussian sigma for velocity smearing is \(0.06\pm0.02 \) in units of \(v/c\), corresponding to a velocity of \(17100^{+4800}_{-4200}   \, \mathrm{km/s}\). The model also gives a significant redshift of  \(z = 0.70^{+0.03}_{-0.02}\), indicating that the line is fitted with a transition from the K-shell of iron ions.

All model parameters and fit statistics are summarized in \autoref{tab:averagespectrum}. The residuals and data-to-model ratios for all four models are presented in \autoref{fig:pl_ratio_tot_rev}  while the unfolded spectra and the residuals for Models 1 and 4 are shown in \autoref{fig:pl_euf_4parti}.

\subsection{Temporal Evolution of the Spectrum}

\begin{figure}[h]
\centering
\includegraphics[width=.45\textwidth]{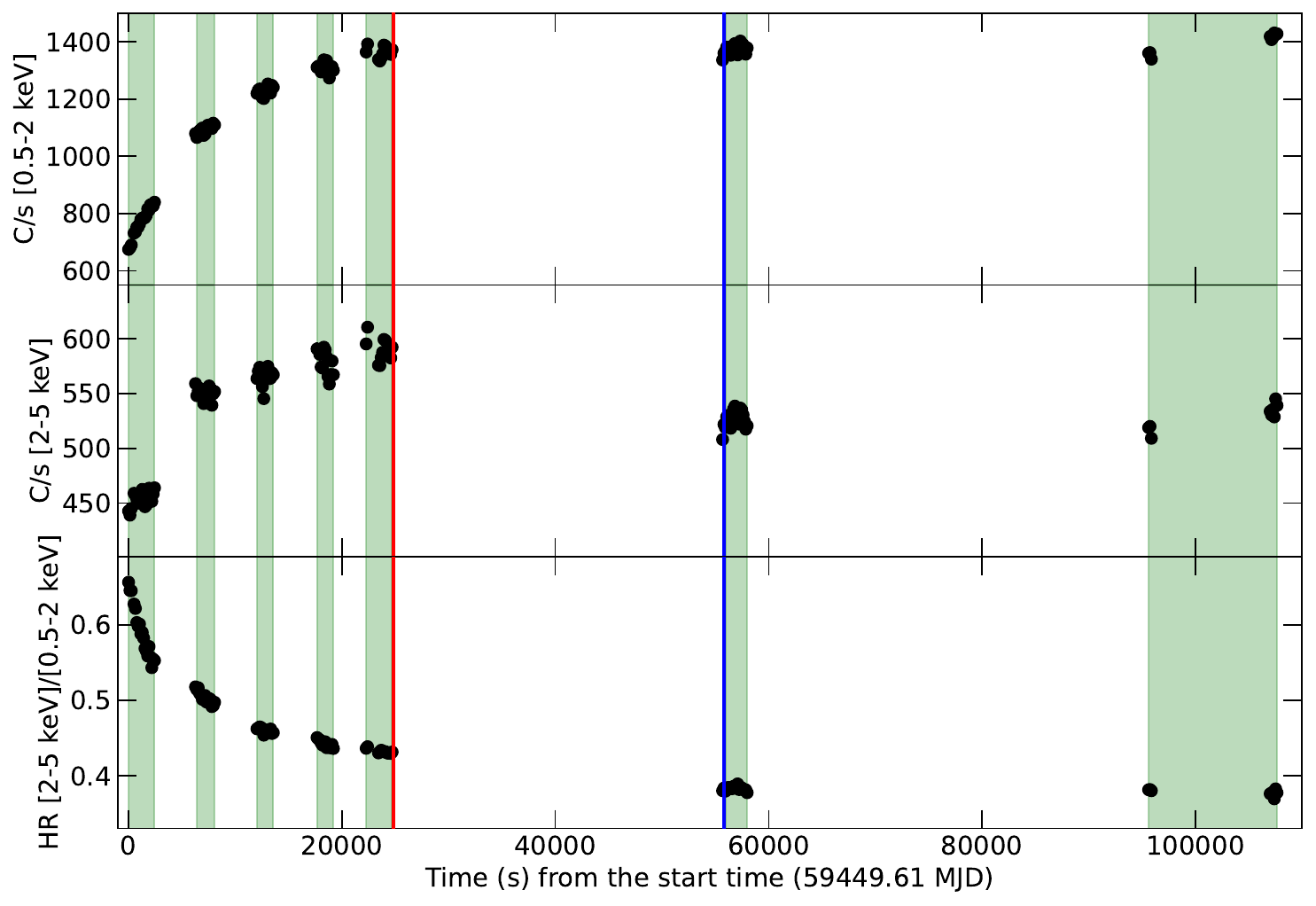}
\caption{\label{fig:lightcurve}
\textit{NICER} light curves for Obs. 105 and 106 in the 0.5–2 keV (top) and 2–5 keV (middle) bands. The bottom panel shows the hardness ratio (HR). Green shaded regions mark intervals used for spectral extraction; red and blue lines indicate the end of Obs. 105 and start of Obs. 106, respectively.}
\end{figure}

\begin{figure}[h]
\centering
\includegraphics[width=0.5\textwidth]{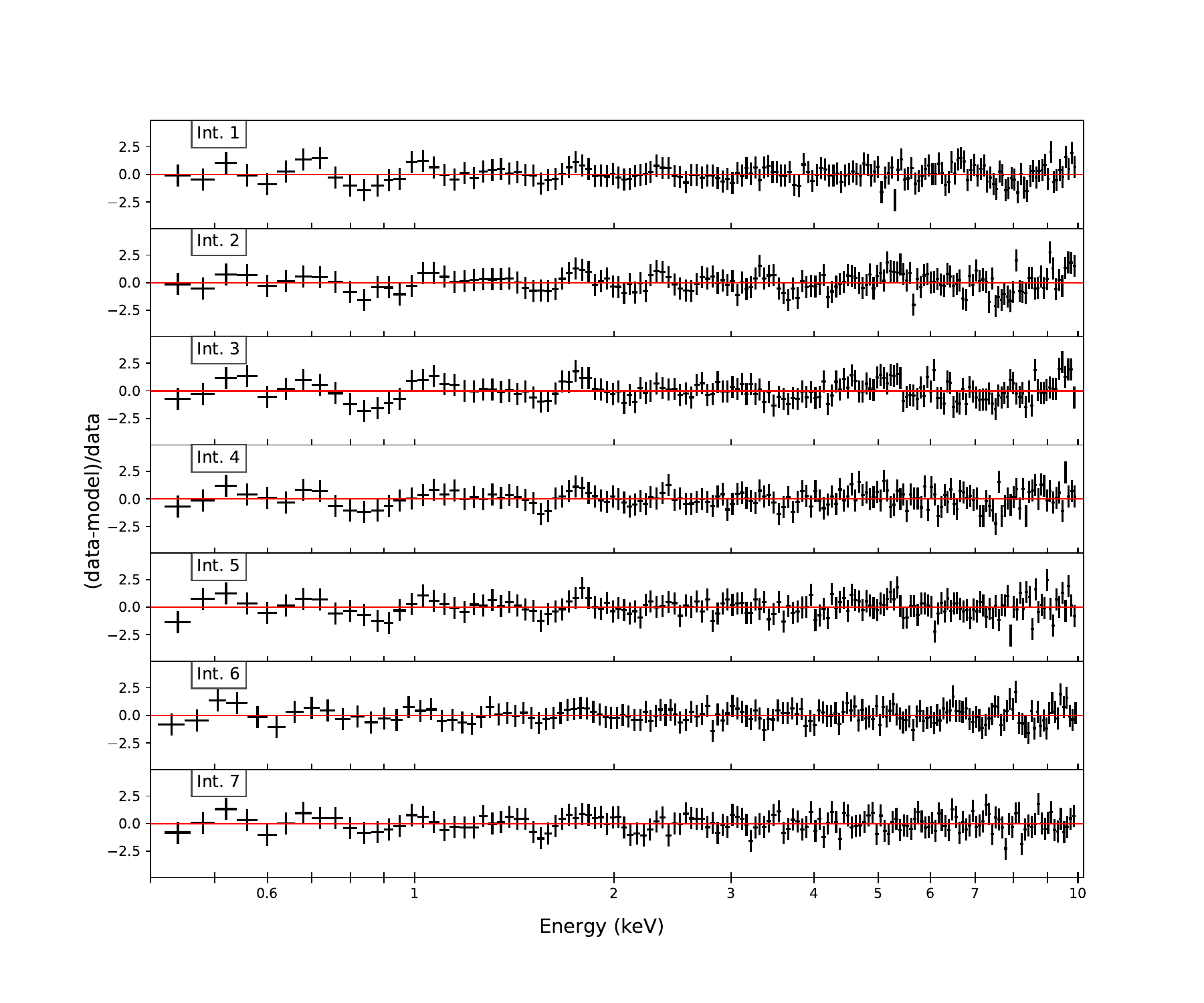}
           \vspace{-20pt} 
\caption{Residuals, expressed in units of $\sigma$, for the 7 spectra extracted from Observations 105 and 106, following temporal segmentation and modeled using \texttt{Model 4}.
}\label{fig:pl_del_intervals}
\end{figure}

\begin{figure}[h]   
\centering
\includegraphics[width=.5\textwidth]{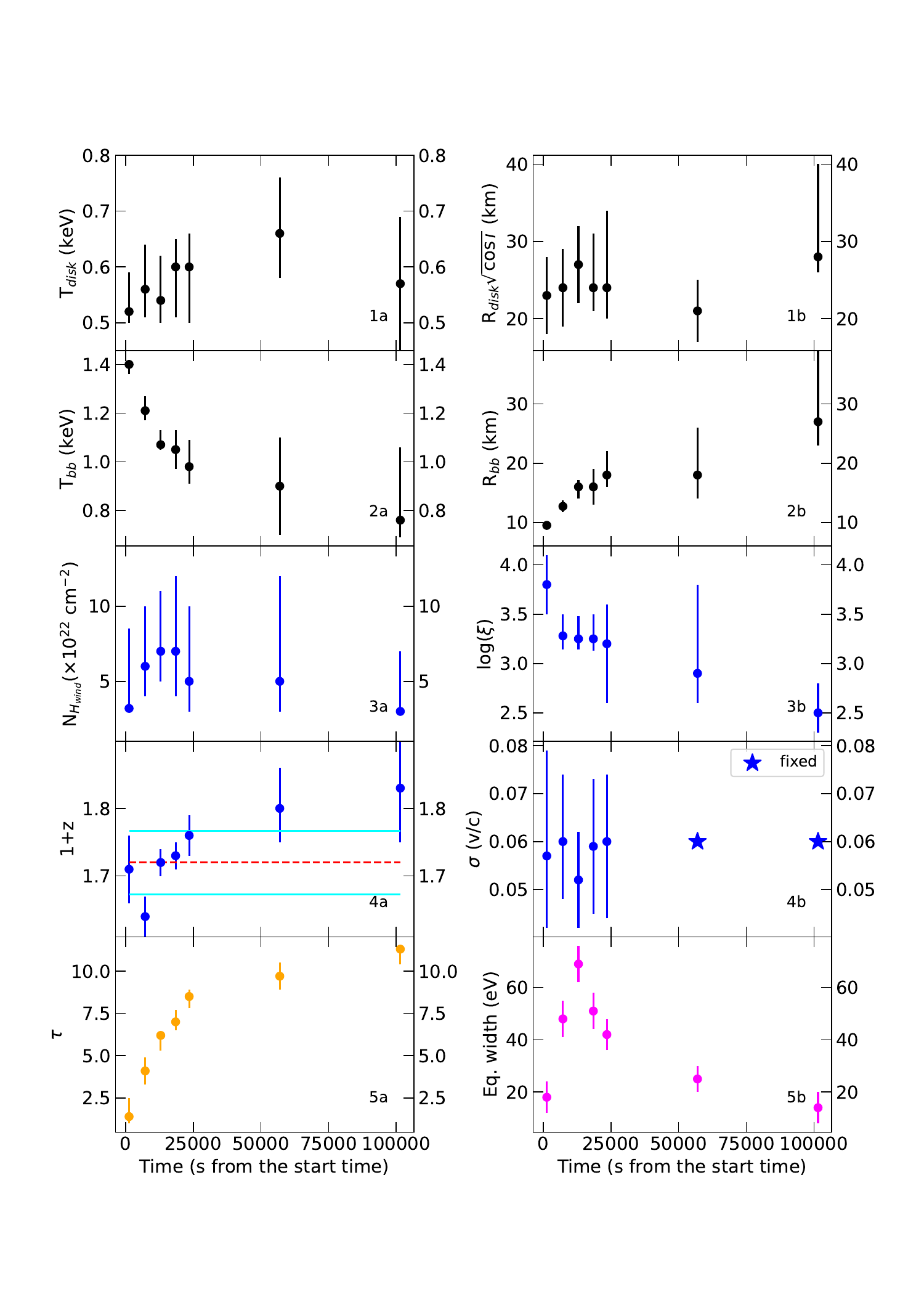}
        \vspace{-30pt} 
    \caption{\label{parameters_plot} Temporal evolution of the best-fit spectral parameters using Model~4. Panels show: inner disk temperature (1a) and radius (1b); blackbody temperature (2a) and radius (2b); Compton cloud optical depth $\tau$ (5a); and, for the {\tt swind1} component, column density (3a), ionization parameter (3b), redshift (4a), and velocity smearing $\sigma$ (4b). Panel 5b shows the equivalent width (in eV) of the broad absorption line derived using Model~3. Errors in all parameters are quoted at the 90\% confidence level, except for the equivalent width in panel 5b, which is shown with a 68\% confidence level.
} 
    \label{fig:trends}
\end{figure}

The hardness ratio (HR) evolution in NICER data reveals significant spectral variability during ObsIDs 105 and 106. To study the spectral evolution, we divided the observations into seven temporal intervals (five for ObsID 105 and two for ObsID 106), as shown in \autoref{fig:lightcurve}. The exposure times of the extracted spectra are listed in \autoref{tab:intervals}. For each segment, we extracted a spectrum and applied Model~4.
The resulting best-fit parameters are listed in \autoref{tab:parameters_7_int_swind}. Residuals are shown in \autoref{fig:pl_del_intervals}, and the evolution of key parameters is illustrated in \autoref{fig:trends}.

The equivalent hydrogen column density of the ionized absorber remains stable within uncertainties. The blackbody temperature ($kT_{\mathrm{bb}}$) shows a monotonic decline from 1.4 keV to 0.8 keV, while its radius ($R_{\mathrm{bb}}$) increases across the observation from 10 km to 30 km. Disk temperature ($kT_{\mathrm{disk}}$) and radius ($R_{\mathrm{disk}} \sqrt{\cos i}$) remain approximately constant at 0.55 keV and  25 km, respectively.

\tabletypesize{\scriptsize}
\begin{deluxetable}{crrr}
\tablecaption{Selected Intervals from the {\it NICER} Observations \label{tab:intervals}}
\tablehead{
\colhead{Interval} & \colhead{Start (s)} & \colhead{Stop (s)} & \colhead{Exp. (s)}
}
\startdata
1 &     0  &   2600  & 2284 \\
2 &  6300  &   8200  & 1716 \\
3 & 12000  &  14000  & 1586 \\
4 & 17500  &  19500  & 1463 \\
5 & 22000  &  25000  & 1454 \\
6 & 55820  &  58120  & 2212 \\
7 & 95000  & 109000  &  897 \\
\enddata
\tablecomments{Times are referenced to the beginning of Obs. 105 (MJD 59449.61).}
\end{deluxetable}

\begin{deluxetable*}{llccccccc}
\tablecaption{Best-fit values of the seven spectra adopting \texttt{Model 4}.\label{tab:parameters_7_int_swind}}
\tablewidth{0pt}
\tabletypesize{\scriptsize}
\tablehead{
\colhead{Component} & \colhead{Parameter} & \colhead{Int. 1} & \colhead{Int. 2} & \colhead{Int. 3} & \colhead{Int. 4} & \colhead{Int. 5} & \colhead{Int. 6} & \colhead{Int. 7}
}
\startdata
\texttt{TBfeo} & $N_H$ ($10^{22}\ \mathrm{cm}^{-2}$) & $0.186\pm0.010$ & $0.188\pm0.007$ & $0.189^{+0.006}_{-0.009}$ & $0.183\pm0.007$ & $0.181^{+0.015}_{-0.005}$ & $0.182^{+0.015}_{-0.007}$ & $0.188^{+0.016}_{-0.005}$ \\
 & $A_O/A_O^{\odot}$ & $1.34\pm0.09$ & $1.27\pm0.08$ & $1.31\pm0.09$ & $1.22\pm0.08$ & $1.22^{+0.15}_{-0.08}$ & $1.28^{+0.10}_{-0.13}$ & $1.29^{+0.10}_{-0.04}$ \\
\texttt{diskbb} & $kT_{\mathrm{disk}}$ (keV) & $0.52^{+0.07}_{-0.02}$ & $0.56^{+0.08}_{-0.05}$ & $0.54^{+0.08}_{-0.04}$ & $0.60^{+0.05}_{-0.09}$ & $0.60^{+0.06}_{-0.10}$ & $0.66^{+0.10}_{-0.08}$ & $0.5\pm0.12$ \\
 & $R_{\mathrm{disk}}\sqrt{\cos{i}}$ (km) & $23\pm5$ & $24\pm5$ & $27\pm5$ & $24^{+7}_{-3}$ & $24^{+10}_{-4}$ & $21\pm4$ & $28^{+12}_{-2}$ \\
\texttt{swind1} & $N_{H_{\mathrm{wind}}}$ & $3^{+5}_{}$ & $5^{+4}_{-2}$ & $7^{+4}_{-2}$ & $7^{+5}_{-3}$ & $5^{+5}_{-2}$ & $5^{+7}_{-2}$ & $3^{+4}_{}$ \\
 & $\log(\xi)$ & $3.8\pm0.3$ & $3.28^{+0.22}_{-0.14}$ & $3.25^{+0.23}_{-0.11}$ & $3.25^{+0.25}_{-0.12}$ & $3.2^{+0.4}_{-0.6}$ & $2.9^{+0.8}_{-0.3}$ & $2.5\pm0.3$ \\
 & $\sigma$ ($v/c$) & $0.057^{+0.022}_{-0.015}$ & $0.060\pm0.014$ & $0.052\pm0.010$ & $0.059\pm0.014$ & $0.06\pm0.02$ & (0.06) & (0.06) \\
 & $z$ & $0.71\pm0.05$ & $0.64\pm0.03$ & $0.72\pm0.02$ & $0.73\pm0.02$ & $0.76\pm0.03$ & $0.80^{+0.06}_{-0.05}$ & $0.83^{+0.17}_{-0.08}$ \\
\texttt{thcomp} & $\Gamma$ & $8^{+2}_{-3}$ & $4.0^{+0.7}_{-0.5}$ & $2.94^{+0.35}_{-0.07}$ & $2.69^{+0.15}_{-0.19}$ & $2.34^{+0.15}_{-0.07}$ & $2.13^{+0.13}_{-0.10}$ & $1.93^{+0.11}_{-0.02}$ \\
 & $kT_e$ (keV) & (3.1) & (3.1) & (3.1) & (3.1) & (3.1) & (3.1) & (3.1) \\
\texttt{bbodyrad} & $kT_{\mathrm{bb}}$ (keV) & $1.40^{+0.02}_{-0.04}$ & $1.21^{+0.06}_{-0.04}$ & $1.07^{+0.06}_{-0.02}$ & $1.05\pm0.08$ & $0.98^{+0.11}_{-0.07}$ & $0.9\pm0.2$ & $0.76^{+0.30}_{-0.07}$ \\
 & $R_{\mathrm{bb}}$ (km) & $9.5\pm0.6$ & $12.7\pm1.0$ & $16.0^{+1.2}_{-2.0}$ & $16\pm3$ & $18^{+4}_{-2}$ & $18^{+8}_{-4}$ & $27^{+24}_{-4}$ \\
\texttt{gaussian} & $I_{\text{\ion{Mg}{12}}}$ ($\times 10^{-2}$) & $0.8\pm0.5$ & $1.5\pm0.5$ & $2.0\pm0.6$ & $1.4\pm0.6$ & $1.1^{+0.8}_{-0.3}$ & $0.6^{+0.6}_{-0.4}$ & $ < 0.6$ \\
\texttt{gaussian} & $I_{\text{\ion{Si}{14}}}$ ($\times 10^{-2}$) & $1.1\pm0.3$ & $2.2^{+0.2}_{-0.4}$ & $2.0\pm0.5$ & $1.5\pm0.4$ & $1.2^{+0.5}_{-0.4}$ & $0.8\pm0.4$ & $< 0.7$ \\
\texttt{gaussian} & $I_{\text{\ion{S}{15}}}$ ($\times 10^{-2}$) & $0.24^{+0.14}_{-0.19}$ & $1.0^{+1.0}_{-0.2}$ & $0.6\pm0.3$ & $0.4\pm0.2$ & $0.3\pm0.3$ &  $< 0.3$ &  $<  0.2$ \\
\texttt{gaussian} & $I_{\text{\ion{Ar}{17}}}$ ($\times 10^{-2}$) & $0.4^{+0.2}_{-0.3}$ & $1.6\pm0.5$ & $1.0\pm0.4$ & $0.7\pm0.5$ & $0.3\pm0.2$ & $ < 0.3 $  & $  <0.2$  \\
$\chi^2$ (dof) &  & 80.7 (144) & 111.1 (143) & 116.6 (143) & 87.5 (143) & 95.3 (143) & 68.8 (146) & 74.7 (138)  
\enddata
\tablecomments{Column densities are in units of $10^{22}$ cm$^{-2}$. Errors are quoted at 90\% confidence. Values in parentheses were kept fixed during fitting. Intensities are given in units of photons cm$^{-2}$ s$^{-1}$. The energies and widths of the emission lines were fixed to the values reported in the text.
}
\end{deluxetable*}
 
\begin{figure}[h]
\centering
\includegraphics[width=.45\textwidth]{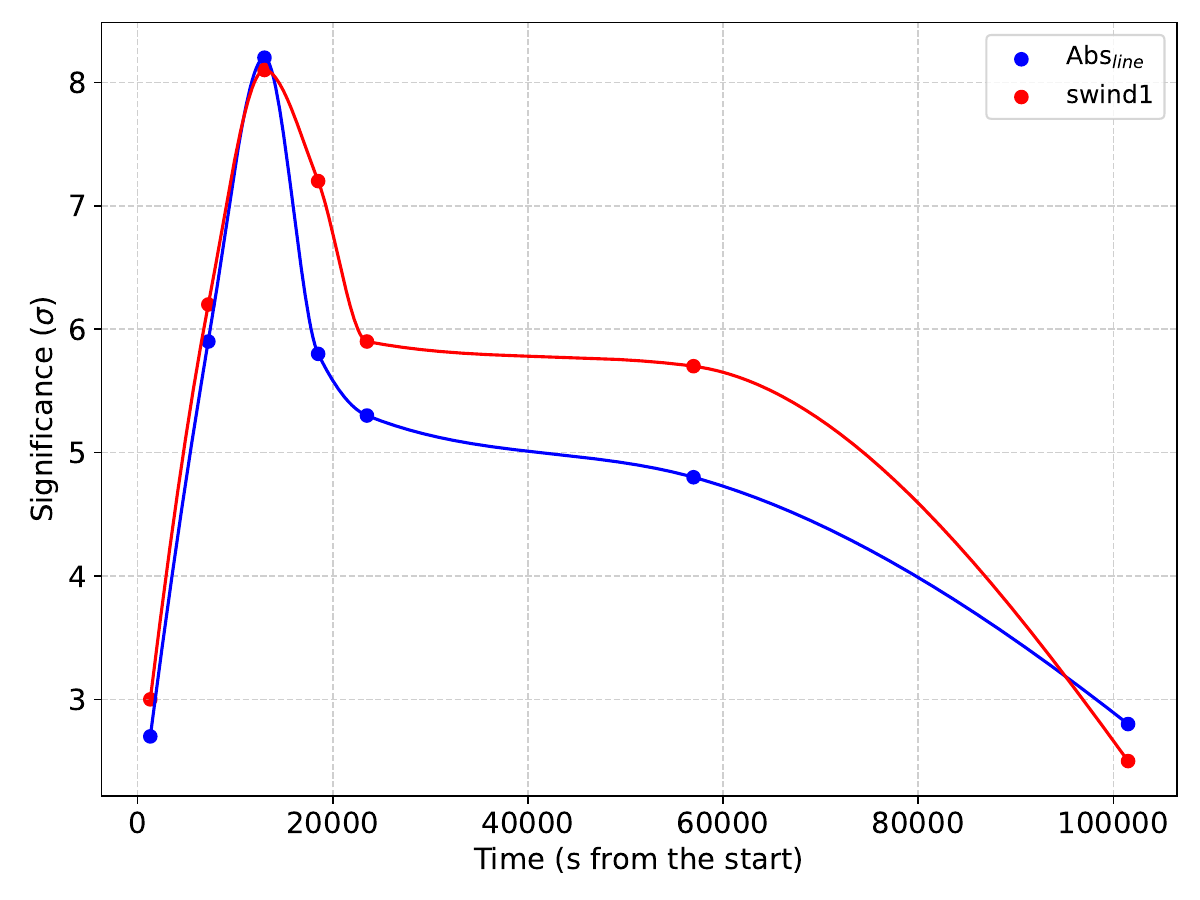}
\caption{The significance, expressed in units of standard deviation, is shown for \texttt{Model 4} (\texttt{swind1} component, red points) and \texttt{Model 3} (\texttt{gaussian} component, blue points), respectively. The absorption feature is barely significant in Interval 7.}\label{fig:significance}
\end{figure}

Each spectrum includes a significant broad absorption feature near 3.8~keV. Its detection significance, assessed by removing the \texttt{swind1} component from the model, ranges from 2.5 to over 8$\sigma$, as shown in \autoref{fig:significance} (red points). The absorption is stronger in intervals 2–4 and becomes marginal in interval 7.

The absorbing ionized matter responsible for the broad absorption line is constrained between \(5 \times 10^{22}~\mathrm{cm}^{-2}\) and \(1.0 \times 10^{23}~\mathrm{cm}^{-2}\) and remains unchanged within the errors throughout the entire observation. 
The ionization parameter, $\log(\xi)$, decreases steadily throughout the observation.
It starts at $3.8 \pm 0.3$ in the first interval, drops to $3.2$ during intervals 2 to 5, decreases further to $2.9$ in interval 6, and finally reaches $2.5$ in interval 7. 

The redshift exhibits a general increase from 0.7 to 0.8, with the exception of interval 2, where it drops to $0.64 \pm 0.03$. For the first five spectra obtained from Obs. 105, the best-fit values are consistent with $z \approx 0.70$, in agreement with the result from the analysis of the averaged spectrum. In contrast, the best-fit values from Obs. 106 are closer to 0.8, albeit with larger associated uncertainties. 
Assuming a constant temporal behavior, we determine a redshift of $1+z = 1.72 \pm 0.05$ at the 99.7\% confidence level.

The optical depth of the Comptonized component, inferred from the parameter $\Gamma$ using Eq.~14 in \cite{2020MNRAS.492.5234Z}, increases progressively from 1.4 to 11.3 throughout the observation.

By adopting Model~3 to fit the seven spectra, we inferred that the equivalent width (in eV) of the broad absorption line varies across the intervals, with values of $18 \pm 6$, $48 \pm 7$, $69 \pm 7$, $51 \pm 7$, $42 \pm 6$, $25 \pm 5$, and $14 \pm 6$ from Interval~1 through Interval~7, respectively (see panel 5b in \autoref{fig:trends}). All quoted uncertainties correspond to a 68\% confidence level. 
The detection significance, evaluated by excluding the Gaussian absorption line from the model, ranges from 2.5$\sigma$ to over 8$\sigma$, as illustrated in \autoref{fig:significance} (blue points). The absorption feature is most prominent in intervals 2–4 and becomes marginal in interval 7, consistent with the trend observed using Model~4.

\section{Constraint on the Photospheric Iron Mass Fraction from the Equivalent Width}

The measurement of the equivalent width (EW) of the broad absorption line provides a quantitative constraint on the mass fraction of iron in the NS photosphere. 
In the linear part of the curve of growth, the rest-frame equivalent width $W_{\lambda,0}$ is related to the ionic column density $N_i$ by the following relation:
\begin{equation}
\label{eq:prima}
W_{\lambda,0} = \frac{\pi e^2}{m_e c^2}\,N_i\,f\,\lambda_0^2= \pi r_e\,N_i\,f\,\lambda_0^2,
\end{equation}
\citep[see e.g.][and references therein]{Iaria2007}, where $e$ is the elementary charge, $m_e$ the electron mass,  $c$ the speed of light,  $r_e$  the classical electron radius, $\lambda_0$ the rest-frame wavelength of the transition, and $f$ the oscillator strength. 
Using the standard values of these quantities gives ${\pi r_e} \simeq 8.85\times10^{-13}$\,{\rm cm}. 
Since ${W_\lambda}/{\lambda} \simeq {W_E}/{E}$, the observed equivalent width can be written as 
$W_{\lambda,\rm obs} = {\lambda_{\rm obs}^2}/{(hc)}\, W_{E,\rm obs}$, 
where $\lambda_{\rm obs}$ is the observed wavelength of the transition and $W_{E,\rm obs}$ the measured EW of the broad absorption line. 
Transforming to the rest frame yields: 
\begin{equation}
\label{eq:seconda}
W_{\lambda,0} = \frac{\lambda_{\rm obs}^2}{hc}\,\frac{W_{E,{\rm obs}}}{1+z}.
\end{equation}
Combining Eqs.~\ref{eq:prima} and~\ref{eq:seconda}, we obtain 
\begin{equation}
\label{eq:terza}
 N_i = \frac{(1+z)\,W_{E,{\rm obs}}}{\pi r_e\, hc\,f},
\end{equation}
where $hc\simeq 1.24 \times 10^{-4}$\,{\rm eV\,cm}. 
Adopting a gravitational redshift of $1+z=1.72$, we find  
$N_i \simeq 1.57  \times 10^{16}\, {W_{E,{\rm obs}}[{\rm eV}]}/{f}$~cm$^{-2}$. 
Using oscillator strengths from \citet{Verner96}, $f=0.798$ for Fe\,XXV and $f=0.416$ for Fe\,XXVI, and an observed EW range of 15--60\,eV, we derive 
$N_i({\rm Fe\,XXV})\simeq(2.9\times10^{17}\text{--}1.2\times10^{18})~{\rm cm^{-2}}$ and 
$N_i({\rm Fe\,XXVI})\simeq(5.7\times10^{17}\text{--}2.3\times10^{18})~{\rm cm^{-2}}$.

Assuming that the line forms over a layer of thickness $L$ comparable to the pressure scale height ($H\simeq 2.5$\,cm at $kT\simeq1$~keV, $g\simeq2\times10^{14}$\,cm\,s$^{-2}$, and a mean molecular weight $\mu=1.3$ for a He-dominated atmosphere, as expected for 4U~1820--303) and adopting an ionization fraction $f_{\rm ion}=0.4$--0.6 consistent with the observed $\log\xi\simeq3.2$--3.8 \citep[see e.g. Fig.~5 in][]{kallman04}, the total iron number density is
\begin{equation}
n_{\rm Fe,tot} = \frac{N_i}{f_{\rm ion} L}.
\end{equation}
The corresponding iron mass fraction is then
\begin{equation}
X_{\rm Fe} = \frac{A m_p N_i}{f_{\rm ion} L \rho_{\rm tot}},
\end{equation}
where the atomic mass number $A$ for iron is 56, $m_p$ is the proton mass, and $\rho_{\rm tot}$ the photospheric density. 
For a helium-rich atmosphere radiating at $F/F_{\rm Edd}\simeq0.2$, appropriate for the post-superburst epoch considered here, we adopt 
$\rho_{\rm tot}\simeq 0.44/(F/F_{\rm Edd})\simeq2$~g\,cm$^{-3}$ \citep[see e.g.][]{suleimanov_11}.

Taking representative values of $W_E=50$\,eV, $L\simeq3$\,cm, and $f_{\rm ion}=0.5$, we obtain 
$X_{\rm Fe} \approx 3\times10^{-5}$ and $X_{\rm Fe} \approx 6\times10^{-5}$ for Fe\,XXV and Fe\,XXVI, respectively. 
Thus, the inferred iron mass fraction at the line-forming depth is 
$X_{\rm Fe} \simeq (3$--$6)\times10^{-5}$, 
consistent (within a factor of 2) with the solar iron abundance $A_{\rm Fe,\odot}=2.7\times10^{-5}$ \citep{Wilms00}.

Finally, the independently fitted warm-absorber column $N_{\rm H,wind}\simeq(0.3$--$1.2)\times10^{23}$\,cm$^{-2}$ 
implies an Fe column of $(0.8$--$3.2)\times10^{18}$\,cm$^{-2}$ for solar composition, in excellent agreement with the ionic columns inferred from the EW analysis. 
We therefore conclude that solar or mildly enhanced iron abundances in the accreted atmosphere are sufficient to reproduce the observed line strength, and that no substantial enrichment from superburst ashes is required.

 \section{Possible Neutron Star Compactness Constraints}

The fitting of the absorption line with Model~4 implies a significant redshift that we interpret as a gravitational redshift.
The  gravitational redshift  is described by the equation 
$$1+z=\left[1-2GM/(c^2R)\right]^{-1/2},$$ 
where $G$ is the gravitational constant, $c$ is the speed of light, $M$ is the NS  mass, and $R$ is the NS radius from which the photons are emitted. The value  of $1+z$ allows us to estimate the NS compactness; by adopting the best-fit  value of $1+z$  of  $1.72 \pm 0.05$ (99.7\% confidence level), we infer a NS compactness of $4.46 \pm 0.13$ km/M\(_{\odot}\)
 (or $3.02 \pm 0.09$ in units of G=c=1).

Recent work by \cite{2024ApJ...975...67J} on thermonuclear X-ray bursts from \xsource, observed with \textit{NICER}, reported a potential spin frequency of \(716 \, \mathrm{Hz}\), with a significance of \(2.9\sigma\). Assuming this spin frequency and an inclination angle of \(31^\circ\) (\citealt{Anitra2025}), we can estimate the NS compactness in its rotating state. 
To achieve this, we consider the Schwarzschild geometry, perturbed to account for a non-zero angular velocity \( \Omega \) while assuming that the object is rigid enough to neglect the quadrupole moment and apply the Newtonian approximation for the angular momentum: \( 2/5 M R^2 \Omega \). 

As rotation breaks spherical symmetry, off-diagonal terms appear in the metric. The relation between the gravitational redshift is then given by
$$  
(1+z)^{-1} = (-g_{tt} - 2 g_{t\phi} \Omega - \Omega^2 g_{\phi\phi})^{1/2}, 
$$
as shown in Eq.~9 of \cite{2010JApA...31..105N}.

Using the perturbative method up to the third order of the Butterworth-Ipser metric \cite{1976ApJ...204..200B} (see also Eq.~33 in \cite{2010JApA...31..105N}) and neglecting the quadrupole moment, we numerically determine the mass as a function of the radius for a neutron star with a spin frequency of \(716 \, \mathrm{Hz}\), an inclination angle of \(31^\circ\), and a gravitational redshift of \(1+z = 1.72 \pm 0.05\).

Fitting our solutions with a quadratic function, we obtain the following relation:
\begin{equation}
\begin{split}
\frac{M}{M_{\odot}} =\ & (-1.64 \pm 0.03)\times 10^{-3} \left(\frac{R}{1\,\mathrm{km}}\right)^2 \\
& + (0.240 \pm 0.007) \left(\frac{R}{1\,\mathrm{km}} \right) - (0.062 \pm 0.001),
\end{split}
\end{equation}
 
\begin{figure}[h]   
\centering 
 \includegraphics[width=.45\textwidth]{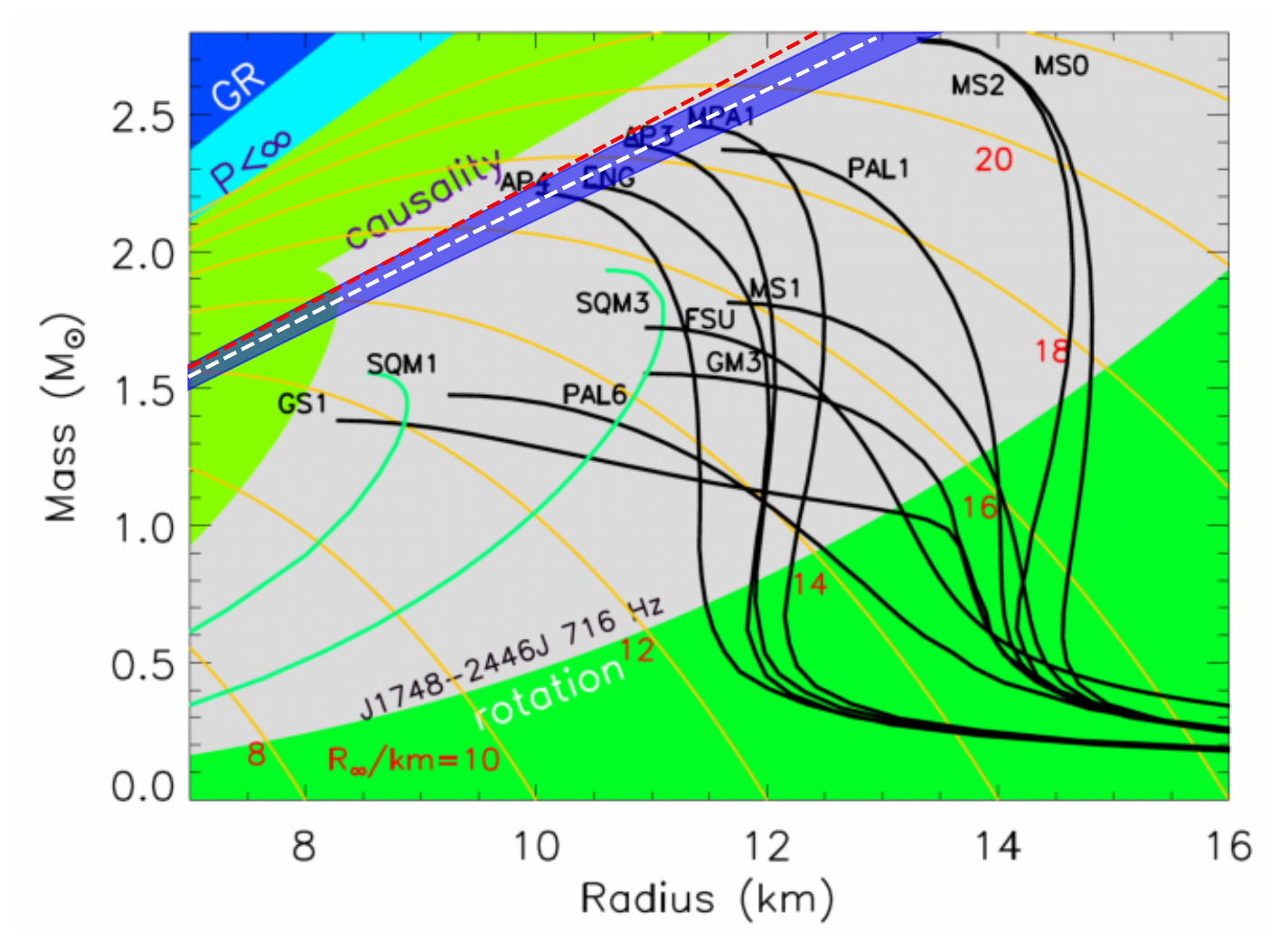}
            \vspace{-0pt} 
    \caption{\label{fig:EOS} 
    Mass-radius diagram for neutron stars. Black (hadronic EOSs) and green (strange quark matter) curves represent common equations of state (see \cite{2001ApJ...550..426L}). The red dashed line shows the best-fit compactness $R/M = 3.02 \pm 0.09$, inferred from a redshift of $1+z = 1.72 \pm 0.05$ (99.7\% confidence) for a non-rotating neutron star. The white dashed line accounts for rotation at 716 Hz. The blue-shaded area reflects the uncertainty in redshift. The green area (upper left) is causally excluded; the lower right is excluded by the fastest known pulsar. Adapted from \cite{2016PhR...621..127L}.} 
\end{figure} 
    
We present the mass-radius relation without accounting for the NS rotation, shown as a dashed red line in \autoref{fig:EOS}. The mass-radius relation indicated by a dashed white line takes into account the proposed NS rotation.  The errors, derived from the uncertainty in the gravitational redshift, are represented by the blue-shaded region.

Adopting the inferred compactness, we can constrain the radius and mass of \xsource, assuming a maximum NS gravitational mass of 2.3 M$_{\odot}$  (\cite{PhysRevD.109.043052}).
The lower limit of the radius is set by the causality condition, which restricts it to values above 8.3 km (see \autoref{fig:EOS}), while the upper limit of 11 km is constrained by assuming a neutron star mass of 2.3 M$_{\odot}$. 
For a radius of 8.3 km, the corresponding lower mass limit is 1.8 M$_{\odot}$. 
Therefore, the NS mass is constrained between 1.8 and 2.3 M$_{\odot}$, while the radius lies within the range of 8.3 to 11 km.

Absorption lines are routinely detected in high-inclination low-mass X-ray binaries \citep[LMXBs; e.g.][]{Boirin2003, Iaria2007,Ueda2004,Diaz2006,Iaria2007b,Ponti2012,Iaria2021}. These features are typically very narrow, with full-width at half maximum (FWHM) less than a few $10^2–10^3$ km s$^{-1}$ (i.e.\ $\Delta E/E < 10^{-3}$, corresponding to $<1–10$ eV at $6–7$ keV), and therefore lie well below the usual spectral resolution of CCD-class X-ray instruments and are often only marginally resolved even with gratings. The lines are widely interpreted as originating in photoionized accretion-disk winds, and their small widths are most likely set by turbulence and/or velocity shear/dispersion within the outflow \citep[e.g.][]{Diaz2016}.
On the other hand, the absorption line discussed here is significantly broad, with a Gaussian $\sigma \simeq 0.25$ keV, compatible with the velocity dispersion inferred from the \texttt{swind1} model of $\sigma / E \simeq 0.06$ in units of $v/c$. We show that this value is compatible even with an NS spin at 716 Hz. In fact, using the best-fit value for the  velocity dispersion inferred from the absorption-line width, \( \sigma = 0.06 \pm 0.02 \, v/c \), we derive a corresponding velocity dispersion expressed as the half-width at half-maximum (HWHM) that is
\( \Delta v \sin{i} = 21,200 \pm 7,000 \, \mathrm{km/s}, \)
where \( i \) is the inclination angle of the source, which is compatible with the broadening that would be induced by the NS rotation if we assume an inclination angle of $31^{\circ}$. Infact, at
the NS surface, the tangential velocity is expected to follow the relation 
\(  v = 2\pi \nu R \), 
where \( \nu \) is the NS spin frequency and \( R \) is the NS radius.  
Considering the extreme values of the estimated NS-radius range, \( R = 8.3 \) km and \( R = 11 \) km, and a NS spin frequency of 716 Hz, we obtain a projected velocity along the line of sight of \( 19,200 \) km/s and \( 25,500 \) km/s, respectively. These values are perfectly consistent with the velocity dispersion inferred from the width of the absorption line.

\section{Discussion}

We report on the study of a pronounced and transient absorption feature in the range of $3.6-3.9$ keV within the \textit{NICER} spectrum of the ultracompact LMXB 4U 1820-30, where a weakly magnetized NS accretes matter from a helium white dwarf companion in an extremely close orbit. This feature emerged in the \textit{NICER} data recorded on August 23, 2021, a few hours after the occurrence of a superburst, and persisted for nearly 60 ks before vanishing.

In the recent literature, the only comparable observation is a transient absorption edge at 3.8~keV identified in the spectrum of the prompt X-ray emission from the gamma-ray burst GRB 990705, as captured by the Gamma-ray Burst Monitor (GRBM) and Wide Field Cameras (WFC) aboard the X-ray satellite BeppoSAX \citep{Amati_2000Sci...290..953A}. This phenomenon was effectively modeled as iron photoelectric absorption by a medium at a redshift of approximately $0.86$, with an iron abundance roughly 75 times that of the solar level.
On the other hand, the absorption feature observed in 4U 1820-30 is not consistent with an absorption edge; instead, it closely fits a Gaussian absorption line profile, suggesting its identification with an iron transition line. 

Absorption lines  similar to that discussed in this paper were  identified in the 1980s with the \textit{Tenma} and \textit{EXOSAT} satellites. Three separate studies \citep[see][] {Waki_1984PASJ...36..819W, Nakamura_1988PASJ...40..209N, Magnier_1989MNRAS.237..729M}, each examining a different LMXB system (4U 1636-536, 4U 1608-522, EXO 1747-214, respectively), reported a pronounced absorption line around 4~keV, interpreted as a 6.7~keV iron line redshifted by the NS gravitational field.
In every instance, the line was detected during type-I bursts, with evidence suggesting a lesser redshift at the outburst's peak—presumably when the NS atmosphere expands due to the thermonuclear explosion. Specifically, in 4U 1636-536 \citep[see][]{Waki_1984PASJ...36..819W}, the line appeared at 5.7 keV at the burst peak, decreasing to 4.1 keV as the burst waned.
These observations imply a high neutron star compactness, yielding an R/M ratio of 3.2. Despite these findings, the existence of these lines and their interpretations have sparked controversy, partly because subsequent observations have not confirmed these claims. 
Moreover, the equivalent width of these absorption lines was of the order of hundreds of eV, which theoretical models of spectral formation in burst atmospheres could not reproduce \citep{Foster_1987MNRAS.228..259F, Day_1992MNRAS.257..471D}. The absorption line we measure in 4U 1820-30 has a moderate and more reasonable equivalent width between 15 and 60 eV.

In all the cases previously mentioned, the line was detected during a type-I burst and, in a few instances, at the burst peak. A more recent study \citep{Barriere_2015} reported a weak absorption line at 5.46 keV at the peak of a type-I burst, adding to the evidence. In contrast, we observe the line well after the occurrence of the superburst.
Fitting the line with a photoionization absorption model (\texttt{swind1} in the \textit{XSPEC} package) 
implies a redshift  of $1.72\pm0.05$ (99.7\% confidence level). Such a redshift is consistent with the line origin from iron ions at the NS surface. 
From the redshift of the line, we can directly estimate the compactness of the star in the range $R/M \simeq 4.46\pm0.13$ km/M$_\odot$, which indicates a mass for the NS larger than $1.8\, M_\odot$ according to the representative modern EOS (see  \autoref{fig:EOS}, adapted from \cite{2016PhR...621..127L}).

The emission profile of a spectral line from the rapidly rotating surface of a NS typically manifests as a single-peak Gaussian under the assumption of uniform emission across the entire star. With an increase in spin frequency, the line tends to weaken and broaden, and its peak emission shifts to higher energies, reflecting the rotational dynamics of the NS \citep{Ozel_2003ApJ...582L..31O}. Line profile asymmetry is accentuated when the emission is restricted to a portion of the stellar surface, a scenario often associated with young NS that exhibit lateral composition and temperature gradients due to intense magnetic fields, leading to an asymmetric, double-peaked profile \citep[e.g.][]{Bhattacharyya_2006ApJ...644.1085B}.
The absorption line observed in \xsource presents a Gaussian profile, suggesting that the emission encompasses the entire (or a significant portion of the) NS surface.

 The obtained compactness was estimated assuming a static neutron star. However, the mass-radius relation must account for rapid rotation if the spin frequency of 716 Hz suggested by \cite{2024ApJ...975...67J} is confirmed. Using a gravitational redshift of $1.72 \pm 0.05$, a spin frequency of 716 Hz, and an inclination angle of $31^\circ$ (\citealt{Anitra2025}), we estimated the mass-radius relation of the neutron star using the Butterworth-Ipser metric \cite{1976ApJ...204..200B}. This approach applies the perturbative method up to the third order, neglecting the quadrupole term and assuming a moment of inertia of $2/5 M R^2$ (see Eq.~33 in \cite{2010JApA...31..105N}).
The mass-radius relation for a rotating neutron star with a spin frequency of 716 Hz is presented as a white dashed line in \autoref{fig:EOS}. The associated uncertainty, driven by the gravitational redshift, is illustrated as the blue-shaded region.

To infer an estimate of the redshift of the absorption line, we applied the \texttt{swind1} model to the Comptonized component.
The best-fit redshift value obtained from this model \(\left(1+z = 1.72 \pm 0.05\right)\) enables us to estimate a neutron star compactness of \(4.46 \pm 0.13\) km/M\(_{\odot}\) or \(3.02 \pm 0.09\) in dimensionless units (\(G=c=1\)).

Using the \texttt{swind1} model, we find that plausible combinations of parameters can reproduce a strong, redshifted iron absorption line without introducing other prominent spectral features, such as an associated Fe edge, that would otherwise be expected.
However, the precise identification of this absorption line requires accurate knowledge of the chemical composition of the neutron star photosphere following a carbon superburst. 
Indeed, alternative identifications have been proposed in the literature. For example, \citet{Peng2025} interpret the line as arising from \ion{Ar}{18} (or \ion{Ca}{20}) originating in the inner accretion disk, while \citet{Jaisawal_2025} suggest that it could be due to neutral Ti or H-like Ca, both cases implying moderate gravitational redshifts.

We note, however, that the NICER observation was performed approximately three hours after the superburst detected by MAXI, and it is therefore reasonable to assume that the wind driven by the burst’s high luminosity had already subsided. In addition, the analysis of the line equivalent width, combined with the inferred absorber column density, indicates iron abundances consistent (within a factor of two) with solar values. Thus, the line does not require any significant enhancement in the abundances of heavy elements. In this scenario, the superburst plays a crucial \emph{indirect} role: the radiation-driven phase associated with the burst removes or substantially thins the accretion corona, temporarily exposing the neutron star photosphere.
As the wind fades and the photosphere returns to its equilibrium radius, the continuum emerging from the stellar surface passes through its own thin, highly ionized atmosphere, allowing the detection of the broad iron absorption line gravitationally redshifted to $\sim$3.8~keV. As the Comptonizing corona gradually reforms around the neutron star, the absorption line becomes progressively weaker, consistent with the decrease in its equivalent width as the optical depth of the corona increases and the ionization parameter of the absorber decreases.

Future instruments with larger collecting areas, such as 
\textit{Athena} and \textit{eXTP} \citep{Barcons_2017AN....338..153B, Zand_2019SCPMA..6229506I, Ray_2018}, which possess the capability for higher throughput and/or improved spectral resolution in the classical X-ray range of $1-10$ keV, could offer new tools for more precise line identifications. These advancements may facilitate the use of such line detections to measure the compactness of neutron stars accurately. Meanwhile, 
\textit{NICER}, thanks to its large effective area in the soft X-ray range, is leading to groundbreaking findings in neutron star physics, particularly through the precise measurement of the mass and radius of neutron stars such as PSR J0030+0451 and PSR J0740+6620. For PSR J0030+0451, two independent analyses using the pulse-profile modeling approach have determined the NS mass to be approximately 1.3 to 1.4 solar masses with a radius of about 13 kilometers \citep{Miller_2019ApJ...887L..24M,Riley_2019ApJ...887L..21R}. For PSR J0740+6620, the inferred gravitational mass of $2.08 \pm 0.07 \, M_\odot$ is the highest reliably determined mass for any NS thus far. The equatorial circumferential radius was found to be around 13.7 km \citep{Miller_2021ApJ...918L..28M}.
In line with that, our results on 4U 1820-30 point towards another massive NS, albeit with a possible smaller radius. Indeed, by using the canonical mass value of $1.4\, M_\odot$ in the compactness we infer from the redshifted absorption line, we obtain a radius for the NS of $6.4 \pm 0.2$ km. Conversely, by setting a radius of 10 km, we derive a mass of $2.18 \pm 0.06\, M_\odot$.
Confirming these results would provide essential constraints on the EoS characterizing ultra-dense matter, further enriching our understanding of the extreme conditions within neutron stars.

\section{Software and third party data repository citations} \label{sec:cite}

This work made use of the following standard software tools and repositories:

\begin{itemize}
  \item \texttt{XSPEC} v12.14.1 for spectral fitting \citep{arnaud1996xspec};
  \item \texttt{HEASoft} v6.34, including \texttt{NICERDAS} and \texttt{FTOOLS}, for data reduction and analysis;
  \item \texttt{CALDB} (NICER calibration database, version xti20240206).
    \item Public NICER data from the HEASARC archive:
  Observation IDs 4050300105 and 4050300106.

\end{itemize}

No proprietary or custom software was used in this work.

\begin{acknowledgments}
R.I. and T.D.S.  acknowledge support from PRIN-INAF 2019 with the project “Probing the geometry of accretion: from theory to observations” (PI: Belloni).
W.L. acknowledges the PhD program in PhD in Space Science and Technology at the University of Trento, Cycle XXXVIII.
A.A. acknowledges financial support from ASI-INAF Accordo Attuativo HERMES Pathfinder operazioni n. 2022-25-HH.0
\end{acknowledgments}

\begin{contribution}
R.I. and T.D.S. led the spectral analysis, data interpretation and manuscript writing.  A.A., F.B.,  A.S. and L. B. contributed to the data interpretation and/or modeling. All authors contributed to the review and editing of the manuscript.

\end{contribution}

\facilities{NICER, MAXI(GSC)}
 
\software{XSPEC \citep{arnaud1996xspec}, HEASoft (\url{https://heasarc.gsfc.nasa.gov/docs/software/heasoft/}), 
NICERDAS (\url{https://heasarc.gsfc.nasa.gov/docs/nicer/analysis_threads/}), 
CALDB (\url{https://heasarc.gsfc.nasa.gov/docs/heasarc/caldb/caldb_intro.html})}


\bibliography{bibliography}{}
\bibliographystyle{aasjournalv7}



\end{document}